# Second-harmonic generation in single-mode integrated waveguides through mode-shape modulation

Jeff Chiles[1,†], Seyfollah Toroghi[1, ‡], Ashutosh Rao[1], Marcin Malinowski[1], Guillermo Fernando Camacho-González[1], Saeed Khan[1], and Sasan Fathpour[1,2,*]

[1]CREOL, The College of Optics and Photonics, University of Central Florida, 4000 Central Florida Blvd., Orlando FL, 32816

[2]Electrical Engineering and Computer Engineering Department, University of Central Florida, 4000 Central Florida Blvd., Orlando FL, 32816

[*]Corresponding author: fathpour@creol.ucf.edu

[†]Now with the National Institute of Standards and Technology (NIST), Boulder, CO

[‡]Now with Partow Technologies LLC, Orlando, FL

A simple and flexible technique for achieving quasi-phase-matching in integrated photonic waveguides without periodic poling is proposed and experimentally demonstrated, referred to as mode-shape-modulation (MSM). It employs a periodic variation of waveguide width to modulate the intensity of the pump wave, effectively suppressing out-of-phase light generation. This technique is applied to the case of second-harmonic generation in thin-film lithium niobate ridge waveguides. MSM waveguides are fabricated and characterized with pulsed-pumping in the near-infrared, showing harmonic generation at a signal wavelength of 784 nm.

The use of integrated photonics for nonlinear frequency conversion is a burgeoning field of research. A critical requirement for achieving coherent frequency conversion is the phase-matching of any involved optical waves. Integrated waveguides can exhibit strong dispersion from the compact geometry and high index contrast, making this more challenging than for bulk optical systems. Several approaches have been investigated for this purpose, such as mode-matching [1-3], microdisk cavities [4] and periodic poling [5]. The former two approaches have the advantage of relatively simple fabrication processes. However, the process of coupling light from the cavities and the use of higher-order modes compromises the conversion efficiency and beam quality of the output.

Periodic poling is a non-resonant approach developed decades ago, which can achieve high conversion efficiencies and maintain excellent beam quality, and since it involves no external cavities, the output light is collected highly efficiently. Unfortunately, it is applicable only to a select few nonlinear optical materials. It is not suitable for GaAs crystals, for instance, which exhibit extremely large nonlinear coefficients (such as the $d_{14}$ element with a value of 90 pm/V) and have a broad transparency range into the far-infrared [6]. Quasi-phase-matching (QPM) in this material system is usually achieved by orientation-patterned substrates [6], which is not readily suitable for sub-micron-scale waveguide integration. Some other means of QPM must then be found to enable integrated and high-performance nonlinear frequency conversion on this and other material systems.

An alternative method of quasi-phase-matching in non-resonant waveguides was proposed by Huang *et al.* in 1998 [7], dubbed "intensity-modulation QPM." In this approach, two parallel waveguides were designed such that the pump light would periodically decouple from that of the second harmonic light being generated. This could be controlled by the gap between the waveguides. In this way, the generation of light would be most efficient when the pump and signal are in-phase, and would be prevented when they are out-of-phase. It places no processing constraints on the material to be used, which is highly valuable. However, the later experimental results were limited by poor mode overlap in the large multimode waveguides employed, and challenging fabrication to achieve the proper coupling coefficients [8].

Considering the practical difficulties with the prior approach, it would be prudent to implement some scheme which does not depend on precisely engineered coupling constants, resonant effects or critical geometrical relations. Accordingly, we propose a novel approach, which will be deemed "mode shape modulation" (MSM).

In this method, an optical waveguide undergoes a periodic variation in its width, the period of which is chosen to compensate the momentum mismatch between the pump and signal waves. As a result of the width variation, the optical mode of the pump wave experiences a modulation of both its size and shape during propagation through the waveguide. As it expands, the peak intensity of the mode decreases. Since the efficiency of the second-harmonic generation process is directly related to the intensity of the pump wave, QPM is achieved by this periodic change in the conversion efficiency. For simplicity, and to avoid optical losses from any abrupt transitions, a sinusoidal width variation pattern has been adopted. This scheme is shown in Fig. 1(a), where a thin slab of material is loaded to become a ridge waveguide with a sinusoidally varying pattern.

Coupled-amplitude equations can be used to describe the process of second harmonic generation in waveguides. The relations are given below [9]:

$$\Delta k = 2k_p - k_s, \quad (1)$$

$$\frac{dA_p}{dz} = -\frac{\alpha_p A_p}{2} + \frac{A_s A_p^* e^{-i\Delta k z} 2i\omega_p^2 d_{eff}}{k_p c^2}, \quad (2)$$

$$\frac{dA_s}{dz} = -\frac{\alpha_s A_s}{2} + \frac{A_p^2 e^{i\Delta k z} 2i\omega_s^2 d_{eff}}{k_s c^2}, \quad (3)$$

where $A$ is the wave amplitude, $z$ is the distance in the propagation axis, $\alpha$ is the propagation loss, $\omega$ is the optical frequency, $d_{eff}$ is the effective nonlinear coefficient in use, $k$ is the angular wavevector, $c$ is the speed of light, and $\Delta k$ is the momentum difference. The subscripts $p$ and $s$ denote applicability to the pump or signal, respectively.

In order to adapt these equations for the case of MSM, the pump's intensity variation must be taken into account. To a first-order approximation, the pump intensity will follow a sinusoidal variation, offset by some average intensity, in good agreement to the geometrical variation imposed on it. This can be modeled by a periodic variation in the power for the pump beam. To conserve optical power in the model, we will introduce a fictitious "reservoir" into which some fraction of the total pump power will oscillate with the period of the MSM grating. Optical power inside the reservoir does not contribute to the nonlinear gain. This fraction corresponds to the amplitude of the intensity variation experienced by the pump beam throughout the grating. We thus modify Eq. 2 to include a reservoir coupling term:

$$\frac{dA_p}{dz} = -\frac{\alpha_p A_p}{2} + \kappa A_r \cos\left(\frac{\pi z}{\Lambda}\right) + \frac{A_s A_p^* e^{-i\Delta kz} 2i\omega_p^2 d_{eff}}{k_p c^2}, \quad (4)$$

and introduce the reservoir amplitude term as

$$\frac{dA_r}{dz} = -\frac{\alpha_p A_r}{2} - \kappa A_p \cos\left(\frac{\pi z}{\Lambda}\right), \quad (5)$$

where $\kappa$ is the coupling coefficient chosen to give the appropriate power coupling ratio, $\Lambda$ is the MSM grating period, and the subscript $r$ denotes the reservoir. The variation in the signal amplitude is not incorporated, as the conversion efficiency does not depend on the signal amplitude in the case of negligible pump depletion.

Next, the simulation parameters were established, with the goal of assessing the performance of MSM. The thin-film LiNbO$_3$ material system (*X*-cut orientation) was selected for this purpose [10-13]. A pump wavelength of 1550 nm was assumed, giving rise to a second harmonic signal at 775 nm. Only the transverse-electric (TE) modes were considered, which results in the d$_{33}$ element of the nonlinear susceptibility tensor (30 pm/V) being used [9]. The waveguides were of a ridge waveguide form, with the ridge element consisting of silicon nitride (SiN), with a refractive index of 1.93 (although the full dispersion behavior is included for all materials). For the MSM structure, a maximum and minimum width of 1.095 and 0.855 µm were used, respectively. The SiN ridge element was 400 nm thick, and the LiNbO$_3$ slab was 600 nm thick. An SiO$_2$ cladding was adopted on top of the waveguide, and the lower cladding also consisted of SiO$_2$, but sufficiently thick to prevent leakage to the substrate underneath. A grating periodicity of 5.52 µm was calculated, which correctly compensates the momentum mismatch between the pump and signal waves, whose effective indices were determined by eigenmode simulations in COMSOL™.

The effectiveness of MSM for QPM depends strongly on the degree of intensity variation that is achieved by the grating, so simulations were conducted to determine the effective mode area at both extremes of the grating width, which determines the fractional change in peak intensity. As seen in Fig. 1(c-d), the mode expands in width when the ridge is narrowed, since the waveguide is being operated in the weak lateral confinement regime. When the ridge is widened, the mode compresses to a smaller area again. The intensity variation is observed to be 10% for this simulation. It is also possible to design an MSM structure in the strong confinement regime (perhaps with wider ridges or a stronger lateral index contrast) to obtain a larger intensity modulation, however, care must be taken to avoid coupling into higher-order modes, which results in loss or deteriorated beam quality. An advantage of weak-confinement MSM structures is their broad single-mode operation; the chosen structure is single-mode at both pump and signal wavelengths in this case.

Next, the intensity of the input light was determined by fixing the input power and dividing by the minimum effective mode area of 1.89 $\mu m^2$. A total power of 100 mW was chosen to be coupled into the simulated waveguide, and loss values of 1 dB/cm were included for both the pump and signal to allow for reasonable grating-induced or material losses. A total propagation length of 10 cm was simulated, but the efficiency figure was chosen from the optimal point along the curve. The initial conditions and coupled amplitude equations (Eqs. 1,3,4,5) were incorporated into an ordinary differential equation solver in MATLAB. The solution is plotted in Fig. 2.

A peak conversion efficiency of 1.9% is achieved at an optimal length of 5.6 cm. At a length of 5 mm, prior to saturation of the signal power, the normalized conversion efficiency is 3.7% $W^{-1}cm^{-2}$. With a larger variation in the intensity modulation, the effect could be enhanced significantly. It should be noted that this is an idealized model, not having taken into account the mode overlap integrals between pump and signal waves. However, a precise model for MSM would need to consider the three-dimensional variation of the overlap integrals due to the different behavior of modes throughout the waveguide grating. This iteration of modeling is more appropriate for future investigation, and is beyond the scope of this proof-of-concept work.

The proposed method of QPM was experimentally investigated on the thin-film LiNbO$_3$ platform (*X*-cut) due to its simplicity of processing, although MSM is suitable for other material systems for future investigations. Waveguides were fabricated on a 4.9-mm-long chip with a slab LiNbO$_3$ thickness of 600 nm, a silicon nitride ridge with a thickness of 400 nm, and a sweep of minimum/maximum widths following a sinusoidal modulation across the direction of propagation.

Various MSM periods were included to provide a range of phase-matching wavelengths. An optical micrograph of the top view of a fabricated MSM waveguide is provided in Fig. 1(b).

The waveguides were characterized by pulsed-pumping with a 500 fs pulse duration source, with a 7 ps pedestal. It is centered at 1560 nm, with an average power of 84 mW. The source spectrum is visible in Fig. 3(a), with an autocorrelation trace in Fig. 3(b). Light was coupled on and off the chip through lensed fibers, with an estimated coupling loss of 6.5 dB/facet. A fiber-based polarization controller was used at the input to align the polarization in the horizontal direction, corresponding to the $Z$-axis of the $LiNbO_3$ film and the TE mode of the waveguide. The output light from the waveguide was collected and then passed into an optical spectrum analyzer (OSA) to determine the phase-matching wavelength for a given MSM period. The best-performing waveguide had a width variation from 855 to 1095 nm, and a periodicity of 5.5 µm, corresponding to an observed phase-matching pump wavelength of 784 nm, in good agreement with that predicted for a similar pump wavelength in the prior simulation. Based on the total insertion loss of 13.5 dB at the pump wavelength, a propagation loss of 1 dB/cm is estimated. The OSA trace showing the generated second-harmonic signal at 784 nm is shown in Fig. 4, with a peak power spectral density (PSD) of -33.3 dB a.u./nm for the signal. The pump shows a PSD of -16 dB a.u./nm at the corresponding phase-matched wavelength of 1568 nm, indicating a penalty of -17.3 dB to the PSD from the SHG process. Differences between the coupling efficiency for the pump and signal waves into the tapered fiber may influence this value slightly.

A plot of integrated pump power vs. integrated output power at the harmonic tone (Fig. 5) confirms the mostly quadratic relationship consistent with the second-harmonic-generation process, showing a slope of 2.18 for a straight-line fit. It should also be noted that the harmonic tone cannot be generated by mode-matching processes, since the waveguide design is single-mode at both the pump and signal wavelengths.

During propagation in the MSM waveguide, it is possible that multiple phase-matched processes may be taking place, which of course includes second-harmonic generation, but possibly also some degree of sum-frequency generation from the broad pump spectrum. As a result, it is not practical to compare the simulation results from earlier with the experimental results, or to report an absolute conversion efficiency, since most of the power of the pump lies outside the useful phase-matching bandwidth of ~ 4 nm. Future investigations with continuous-wave input will allow a more precise, normalized conversion efficiency to be measured. The conversion efficiency could be substantially increased by employing a higher-refractive-index loading material to allow faster compression of the mode. Additionally, there may be optimized, non-sinusoidal width modulation patterns which achieve low grating-induced losses while maintaining strong intensity modulation.

In conclusion, a flexible means of achieving quasi-phase-matching in integrated photonic wavelength converters has been proposed, simulated and experimentally demonstrated. This method, called mode-shape-modulation (MSM), works by imposing a sinusoidal variation in waveguide width along the propagation axis of a nonlinear waveguide. It has advantages over prior integrated QPM approaches in its fabrication tolerances and applicability to many waveguide geometries. Additionally, it preserves a high beam quality for the output light by operating in single-mode waveguides. MSM devices were fabricated and characterized by near-infrared pulsed pumping. Second-harmonic generation was observed at a signal wavelength of 784 nm. This opens up the possibility of integrated second-order-nonlinear frequency converters on other material platforms such as aluminum gallium arsenide (AlGaAs) and indium gallium phosphide (InGaP) that are incompatible with conventional periodic poling techniques.

This project is being supported by the U.S. Office of Naval Research (ONR) Young Investigator Program and the U.S. Defense Advanced Research Projects Agency (DARPA). The views, opinions and/or findings expressed are those of the author and should not be interpreted as representing the official views or policies of the Department of Defense or the U.S. Government.

**Figures**

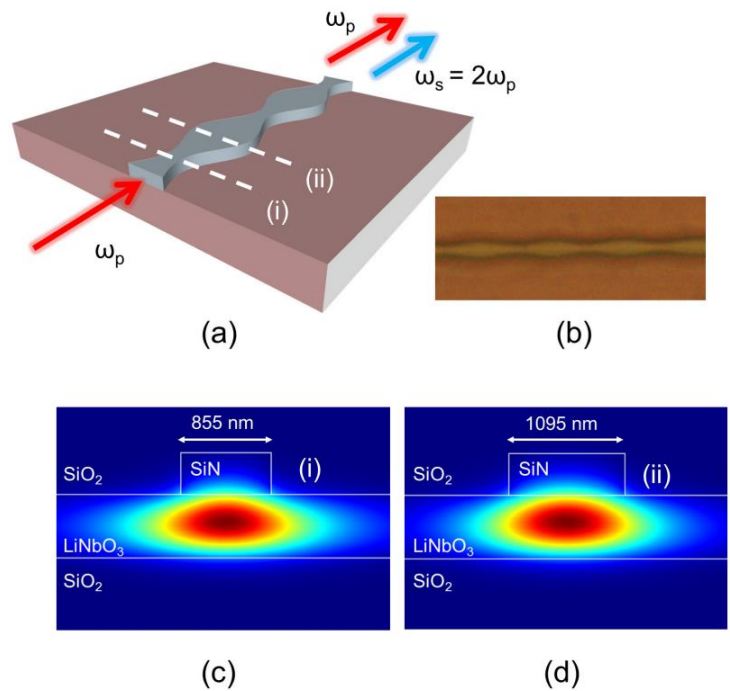

(a)                                                 (b)

(c)                                                 (d)

**Fig. 1**. (a) Concept of the MSM ridge waveguide, showing the longitudinally varied waveguide width following a sinusoidal pattern (exaggerated in its magnitude for visibility); (b) top-view optical micrograph of a fabricated MSM ridge waveguide; and intensity profile of the fundamental TE mode at cross-sections (c) *i* and (d) *ii*.

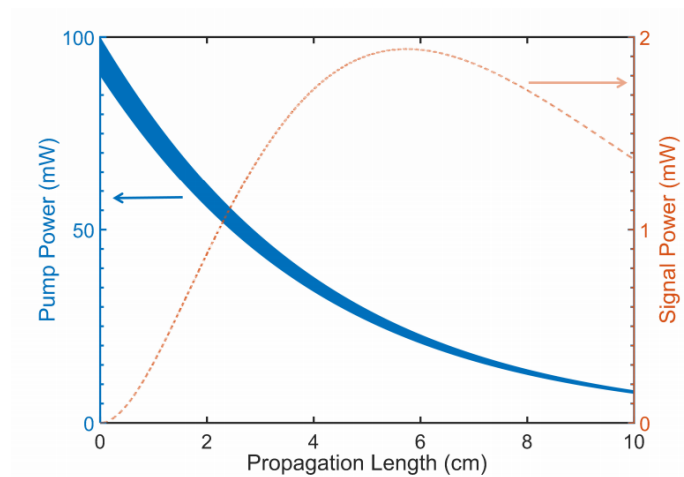

**Fig. 2**. Conversion efficiency versus propagation length for the MSM numerical simulation. Left axis (blue line): Pump power coupled into the waveguide. The thickness of the line represents the oscillations in intensity due to the MSM grating. Right axis (dashed red line): Signal or second-harmonic power generated.

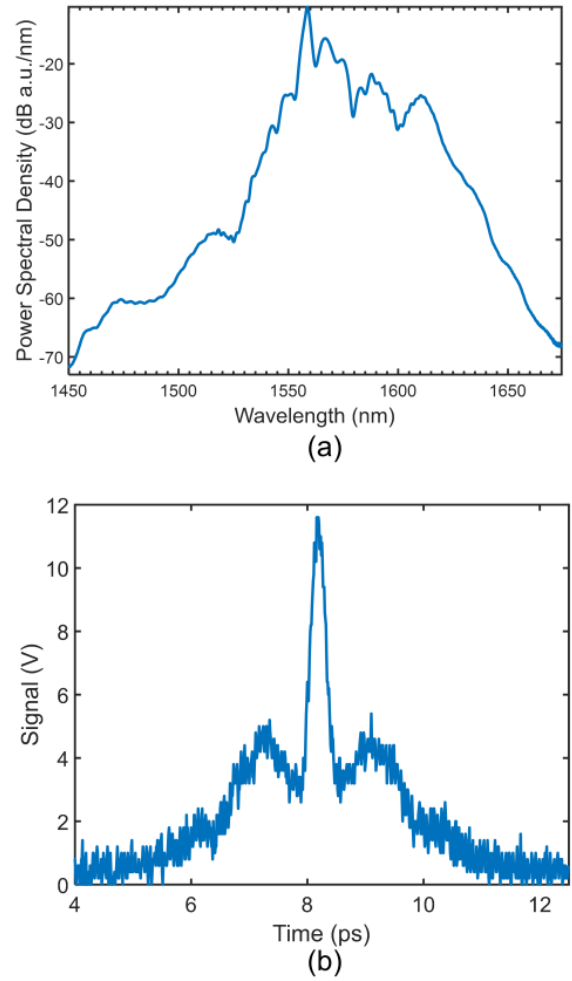

**Fig. 3.** Pulsed-pump characteristics: (a) power spectrum captured on an OSA trace; (b) autocorrelation trace.

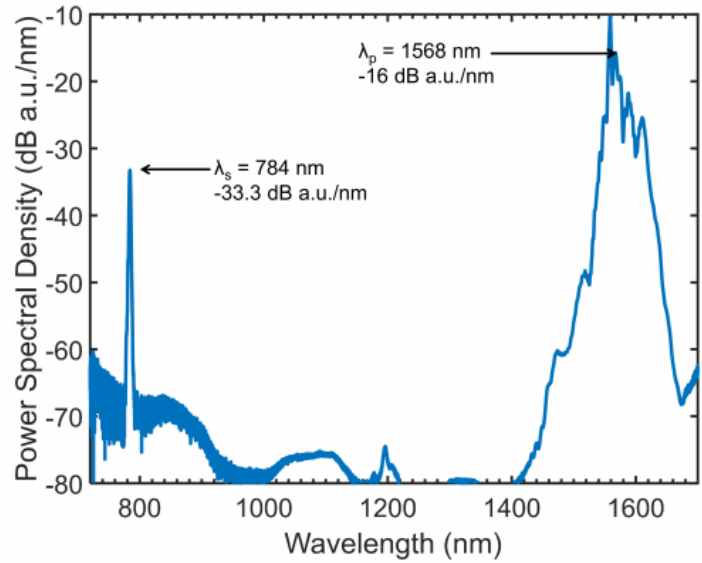

**Fig. 4.** OSA trace of the output power spectral density, showing second-harmonic light being generated at a wavelength of 784 nm.

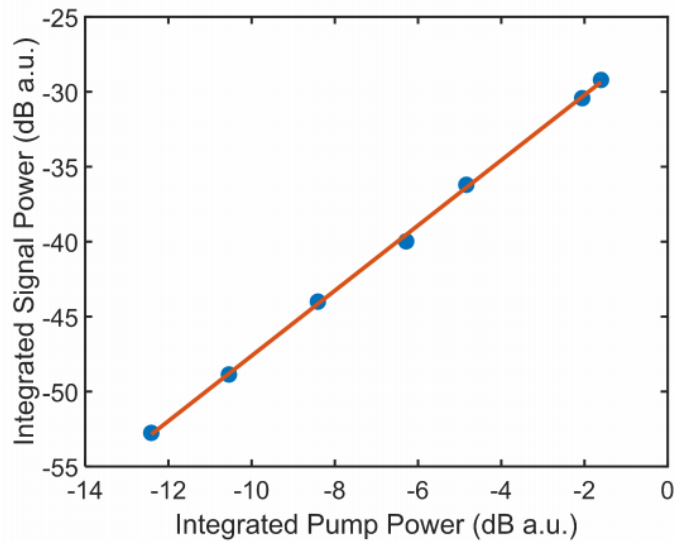

**Fig. 5.** Pump power vs. signal power (dots), showing the slope of 2.18 (fitted line).